\documentclass[pra,twocolumn,aps,nopacs]{revtex4}

\usepackage{enumerate}
\usepackage{amsmath}
\usepackage{amssymb}
\usepackage{mathtools}
\usepackage{quantum}
\usepackage{color}
\usepackage[colorlinks,citecolor=blue,linkcolor=blue,linktocpage,breaklinks,hypertexnames=false,pdfpagelabels]{hyperref}
\usepackage[amsmath,thmmarks,hyperref]{ntheorem}
\usepackage[capitalise]{cleveref}

\usepackage{psfrag}
\usepackage[dvips]{graphicx}
\usepackage{enumitem}

\crefname{condition}{Condition}{Conditions}
\creflabelformat{condition}{\textup{(#2#1#3)}}

\newtheorem{theorem}{Theorem}

\newtheorem{result}[theorem]{Result}

\theorembodyfont{\normalfont}

\theorembodyfont{\normalfont}
\newtheorem{remark}[theorem]{Remark}

\theoremstyle{break}
\newtheorem{theorem-break}[theorem]{Theorem}
\newtheorem{lemma-break}[theorem]{Lemma}
\newtheorem{corollary-break}[theorem]{Corollary}
\newtheorem{definition-break}[theorem]{Definition}

\theoremstyle{nonumberplain}
\theorembodyfont{\normalfont}

\theoremsymbol{$\Box$}
\newtheorem{proof}{Proof}

\makeatletter
\def\p@subsection{}
\def\p@subsubsection{}
\makeatother

\newcommand{\ra}{\rangle}
\newcommand{\la}{\langle}
\newcommand{\beq}{\begin{equation}}
\newcommand{\eeq}{\end{equation}}
\newcommand{\be}{\begin{eqnarray}}
\newcommand{\ee}{\end{eqnarray}}
\newcommand{\ba}{\begin{array}}
\newcommand{\ea}{\end{array}}
\newcommand{\bc}{\begin{center}}
\newcommand{\ec}{\end{center}}
\newcommand{\ben}{\begin{enumerate}}
\newcommand{\een}{\end{enumerate}}
\newcommand{\bi}{\begin{itemize}}
\newcommand{\ei}{\end{itemize}}
\newcommand{\bt}{\begin{table}}
\newcommand{\et}{\end{table}}
\newcommand{\btab}{\begin{tabular}}
\newcommand{\etab}{\end{tabular}}
\newcommand{\bfi}{\begin{figure}}
\newcommand{\efi}{\end{figure}}
\newcommand{\im}{\item}
\newcommand{\bd}{\begin{description}}
\newcommand{\ed}{\end{description}}

\newcommand{\nn}{\nonumber}

\newcommand{\mc}{\mathcal}

\newcommand{\trm}{\textrm}

\newcommand{\lb}{\label}
\newcommand{\eref}[1]{(\ref{#1})}

\newcommand{\rk}{\textrm{rank}}

\begin{document}

\title{Purifications of multipartite states: limitations and constructive methods}

\author{Gemma De las Cuevas$^{1}$, Norbert Schuch$^{2}$, David P\'erez-Garc\'ia$^{3}$, and J.~Ignacio Cirac$^{1}$}

\affiliation{$^{1}$Max Planck Institute for Quantum Optics, Hans-Kopfermann-Str.~1,  D-85748 Garching, Germany\\
  $^{2}$ Institut f\"ur Quanteninformation, RWTH Aachen, Aachen, Germany\\
$^{3}$Departamento de An\'{a}lisis Matem\'{a}tico, Facultad de CC Matem\'{a}ticas, Universidad Complutense de Madrid, 28040 Madrid, Spain}

\begin{abstract}
We analyze the description of quantum many-body mixed states using matrix product states and operators.
We consider two such descriptions: 
(i) as a matrix product density operator of bond dimension $D$, and 
(ii) as a purification that is written as a matrix product state of bond dimension $D'$.  
We show that these descriptions are inequivalent in the sense that 
$D'$ cannot be upper bounded by $D$ only. 
Then we provide two constructive methods to obtain (ii) out of (i).
The sum of squares (sos) polynomial method scales exponentially in the number of different eigenvalues,  
and its approximate version is formulated as a Semidefinite Program, which gives efficient approximate purifications whose $D'$ only depends on $D$. 
The eigenbasis method scales quadratically in the number of eigenvalues, 
 and its approximate version is very efficient for rapidly decaying distributions of eigenvalues. 
Our results imply that a description of mixed states which is both efficient and locally positive semidefinite does not exist, but that good approximations do.   
\end{abstract}

\date{\today}

\maketitle

%%===================================
\section{Introduction}

Quantum many-body systems appear in a variety of fields in physics, such as condensed matter, quantum chemistry, or high-energy physics. 
Since their Hilbert space description is 
intractable (as it scales exponentially with the system size), a number of methods have been proposed to describe them efficiently. 
One of them are tensor networks \cite{Ha06b}, which have been particularly successful in describing one-dimensional pure states  with Matrix Product States (MPS)  \cite{Fa92}.
%----BEGIN CHANGE 2013-10-----
Its canonical form has facilitated the distinction between injective and non-injective MPS \cite{Pe07}, 
(which determines the ground state degeneracy of the parent Hamiltonian, and is linked to many other physical properties,  see e.g.~\cite{Ci13,Wa13,Wa12b}), which has led to the classification of gapped phases of one-dimensional systems \cite{Sc11}. 
This mathematical understanding has allowed to characterize global properties of the state, such as topological order or symmetries, in a local way \cite{Pe08,Sc10b}. 
This is in contrast with one-dimensional \emph{mixed states}, whose description with tensor networks is much more scarce.
%----END CHANGE 2013-10-----
While the class of Matrix Product Density Operators (MPDOs) has been defined \cite{Ve04d,Zw04},  a canonical form has not been found. 
This is despite the fact that mixed states are used to describe quantum many-body systems at finite temperature, systems out of equilibrium, dissipative dynamics, or  lack of knowledge of the state of the system.

\begin{figure}[thb]\centering
\psfrag{r}{$D$}
\psfrag{s}{$D$}
\psfrag{p}{$D'$}
\psfrag{q}{$D'$}
\psfrag{O}{$M_{1}$}
\psfrag{P}{$M_{2}$}
\psfrag{Q}{$M_{3}$}
\psfrag{R}{$M_{N}$}
\psfrag{A}{$A_{1}$}
\psfrag{B}{$A_{2}$}
\psfrag{C}{$A_{3}$}
\psfrag{D}{$A_{N}$}
\psfrag{a}{$\bar A_{1}$}
\psfrag{b}{$\bar A_{2}$}
\psfrag{c}{$\bar A_{3}$}
\psfrag{d}{$\bar A_{N}$}
\psfrag{1}{(a)}
\psfrag{2}{(b)}
\psfrag{?}{?}
\includegraphics[width=0.48\textwidth]{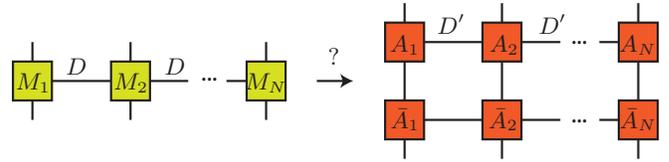}
\caption{(Left) The MPDO form of a mixed state writes it as a sum of $D$ tensor products. The matrices $M_{i}$ are generally not positive semidefinite. 
(Right) The local purification form of $\rho$, where the purifying state is written as an MPS of bond dimension $D'$.
The question is: can $D'$ be upper bounded by $D$?
}  
\label{fig:problem} 
\end{figure}

One of the difficulties in defining a canonical form for mixed states is that with MPDOs one cannot verify locally that the global tensor is positive semidefinite.
This implies that local truncations of the tensors generally destroy the global positivity, 
causing instability in numerical algorithms. 
An alternative is to use the local purification form, where the mixed state is purified and the purification is expressed as an MPS. 
While its local tensors are positive semidefinite, it is not known if an efficient MPDO form guarantees an efficient local purification. 
That is, are the two descriptions equivalent, or can the local purification be much more costly than the MPDO form (see figure~\ref{fig:problem})?

In this paper we address this question and show that these two descriptions are \emph{inequivalent}. 
Specifically, we provide classical multipartite states whose MPDO form has a fixed cost, but whose local purification form has an unbounded cost. 
Then, we provide \emph{two constructive purification methods}  applicable to all multipartite density matrices, which relate the two forms but also involve the number of (different) eigenvalues. 
The sum of squares (sos) polynomial method has an exact version which scales exponentially in the number of different eigenvalues. 
Its approximate version is formulated as a Semidefinite Program (SDP),
which shows an efficient and robust behavior for all the tested distributions of eigenvalues. 
The eigenbasis method has an exact version which scales multiplicatively with the number of eigenvalues, and its approximate version gives very efficient purifications for rapidly decaying distributions of eigenvalues. 

This paper is organized as follows. 
First, we present the problem in Sec.~\ref{sec:setting}. 
Then we show the inequivalence of the MPDO form and the local purification form in Sec.~\ref{sec:counterexample}.
In Sec.~\ref{sec:purif} we present the two purification methods: 
the sos polynomial method (Sec.~\ref{ssec:1method}), and 
the eigenbasis method (Sec.~\ref{ssec:2method}), both with its main idea, its exact and its approximate version, and in Sec.~\ref{ssec:compare} we compare the two approximate methods. 
Finally we conclude and mention further directions in Sec.~\ref{sec:conclusions}.

\section{The setting}
\lb{sec:setting}

We now present the question that concerns us, 
which is whether the MPDO form and the local purification form of a mixed state can be related. 
We will first present some notation and definitions (Sec.~\ref{ssec:def}),
and then introduce the problem  (Sec.~\ref{ssec:problem}).

\subsection{Definitions}
\lb{ssec:def}

Let us first fix some notation. 
We write $A \succeq 0$  to denote that the matrix $A$ is positive semidefinite (i.e.~Hermitian with nonnegative eigenvalues), and  $A\geq 0$ to denote that  it is nonnegative (i.e.~with nonnegative entries).
We also write $\trm{rank}(\rho)$ ($\trm{rank}(A)$) to denote the number of non-zero eigenvalues (non-zero singular values) of the matrix $\rho$ ($A$).
The trace norm of a matrix $A$ is defined as  $||A||_{1}=\sum_{i=1}^{r} s_{i}$, 
where $r=\trm{rank}(A)$ and $s_{i}$ are its singular values.  
Given a multipartite state with $N$ subsystems arranged on a one-dimensional line (henceforth called a 1D state),
we call a linear bipartition a splitting of the form $1\ldots k$ vs. $k+1,\ldots, N$ for any $k$. 
Finally, given a multipartite state $|\psi\ra=\sum \psi_{i_{1}\ldots i_{N}}|i_{1}\ldots i_{N}\ra$ with Schmidt rank $D'_{k}$ across the bipartition $i_{1}\ldots i_{k}$ vs.~the rest, we define the Schmidt rank of $|\psi\ra$ as   $\trm{SR}(\psi)=\max_{k}D'_{k}$.

We focus on 1D mixed states with $N$ $d$-level systems and open boundary conditions, described by a density matrix $\rho$,
\begin{eqnarray}
\rho=\sum_{i_{1},j_{1},\ldots,i_{N},j_{N}=1}^{d} 
\rho_{i_{1},\ldots,i_{N}}^{ j_{1},\ldots,j_{N}} \:
|i_{1},\ldots,i_{N}\ra\la j_{1},\ldots,j_{N}|\, .
\lb{eq:rho}
\end{eqnarray}
The are two natural ways of describing $\rho$ locally. The first one is \emph{MPDO form}\cite{Ve04d,Zw04}, which   is defined as
\be
\rho= \sum_{\alpha_{1}=1}^{D_{1}} \ldots \sum_{\alpha_{N-1}=1}^{D_{N-1}}
M_{1}^{\alpha_{1}}\otimes M_{2}^{\alpha_{1},\alpha_{2}}\otimes \ldots
\otimes M_{N}^{\alpha_{N-1}}\ ,
\lb{eq:osd}
\ee
where $M_{k}^{\alpha_{k-1},\alpha_{k}}$ are $d\times d$ matrices for $1< k<N$,
and $M_{1}^{\alpha_{1}}$ ($M_{N}^{\alpha_{N-1}}$) is a row (column) vector of size $d$.
Note that the subindex only indicates the subsystem that the matrix is describing.
Here $D_{k}$ (for all $k$) is the minimal dimension such that \eref{eq:osd} holds
\footnote{The minimal $D_{k}$ (for all $k$) is unique as it does not depend on the order on which the decompositions are made. To see this, note that $D_{k}$ is the bond dimension of the bipartition $i_{1}j_{1}\ldots i_{k}j_{k}$ vs.~the rest of the pure state $|\psi\ra= \sum\rho_{i_{1},\ldots,i_{N}}^{ j_{1},\ldots,j_{N}} |i_{1},\ldots,i_{N}\ra| j_{1},\ldots,j_{N}\ra $.}. 
The \emph{operator Schmidt rank} of $\rho$ is defined as 
\footnote{The operator Schmidt rank of $\rho$ must not be confused with its tensor rank, which is the minimal $r$ such that $\rho=\sum_{\alpha=1}^{r} M_1^{\alpha} \otimes M_2^{\alpha} \otimes \ldots \otimes M_{N}^{\alpha}$.} % It is NP-hard to determine the tensor rank of a state.
\be
\trm{OSR}(\rho):=\max_{k}D_{k}=D\, .
\ee

The second form is the \emph{local purification form}, which  is obtained by purifying the mixed state $\rho$ (living in system $S$) into a pure state $|\Psi\ra$ (living in $S,S'$), and expressing $|\Psi\ra$ as an MPS
\footnote{There are infinitely many purifications of a given mixed state, and they are all related by an isometry on the ancillary system. We always consider the purification which minimizes the Schmidt rank of the purifying state along every linear  bipartition and which is local (i.e.~an ancillary subsystem $S'_{i}$ is attached to each subsystem $S_{i}$). },
\be
&&\rho=\tr_{S'}|\Psi\ra\la\Psi|\\ 
&&|\Psi\ra=
\sum_{\beta_{1}=1}^{D'_{1}} \ldots \sum_{\beta_{N-1}=1}^{D'_{N-1}}
A_{1}^{\beta_{1}} \otimes A_{2}^{\beta_{1},\beta_{2}} \ldots A_{N}^{\beta_{N-1}}\, .
\lb{eq:purif}
\ee
Here 
$A_{k}^{\beta_{k-1},\beta_{k}}$ are $d\times d_{a_{k}}$ matrices for $1<k<N$, where $d_{a_{k}}$ is the dimension of the local ancilla, and $A_{1}^{\beta_{1}}$ ($A_{N}^{\beta_{N-1}}$) is a row (column) vector of size $d\times d_{a_{1}}$ ($d\times d_{a_{N-1}}$).
Here $D'_{k}$ is the Schmidt rank of $|\Psi\ra$ of the bipartition $\tr_{k+1\ldots N}|\Psi\ra\la\Psi|$ vs.~the rest.
We define the \emph{purification rank} of $\rho$ as 
\be
\rk_{\trm{puri}} (\rho):=\max_{k} D'_{k}=D'\, .
\ee

\subsection{The problem}
\lb{ssec:problem}

We want to find out if the MPDO and the local purification form are equivalent, or if the latter can be arbitrarily more costly than the former. 
The advantage of the local purification form is that the local tensors are positive semidefinite, since they are of the form $A_{k}A_{k}^{\dagger}$.
In contrast, in the MPDO form this is generally not true, i.e.~$M_{k}^{\alpha_{k-1},\alpha_{k}}\nsucceq 0$, thus it is locally invisible that $\rho \succeq 0$.
This is a problem theoretically as well as numerically, as local truncations destroy the global positivity. 

\begin{figure}[thb]\centering
\psfrag{1}{$D\!=\!2$}
\psfrag{2}{$D\!=\!1$}
\psfrag{3}{$D'\!=\!2$}
\psfrag{4}{$D'\!=\!1$}
\psfrag{5}{$D\!=\!D'\!=\!1$}
\psfrag{H}{$\mc{H}$}
\includegraphics[width=0.48\textwidth]{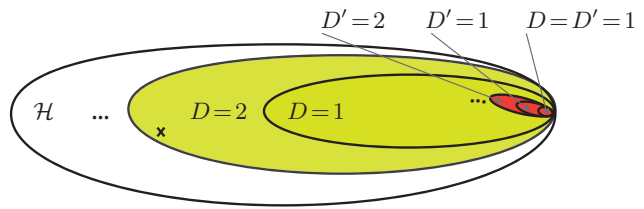}
\caption{The Hilbert space  $\mc{H}$ of mixed states. 
How much larger is the family of states with fixed operator Schmidt rank $D$ compared to that with fixed purification rank $D'$? 
Are they comparable in size, 
or are there states requiring very small $D$ and very large $D'$ (such as the one marked with a cross)?
}  
\label{fig:families} 
\end{figure}

On the other hand, the problem of the local purification form is that it is not known how much larger the purification rank $D'$ may be compared to the operator Schmidt rank $D$ (see figure~\ref{fig:families}). 
Thus, we want to see if there is a transformation from the MPDO to the local purification form in which the bond dimension increases in a controlled way.
That is, we want to find out if $D'$ can be upper bounded by a function of $D$.  

Observe that it is very easy to obtain a bound in the other direction, namely $D\leq D'^{2}$, simply by choosing
\be
[M_{k}]_{i_{k},j_{k}}
=\sum_{z_{k}=1}^{d_{a_{k}}}
[A_{k}]_{i_{k},z_{k}} \otimes ([A_{k}]_{j_{k},z_{k}})^{\dagger}\, ,
\ee 
where $i_{k},j_{k},z_{k}$ are physical indices. 
On the other hand, both ranks can be upper bounded by the physical dimension, 
\be
D\leq d^{N}\, ,\qquad D' \leq d^{N} \,. 
\ee 

\section{Inequivalence of the two forms}
\lb{sec:counterexample}

We now show that the MPDO form and the local purification form are inequivalent.
More precisely, we provide a family of multipartite classical states with a constant operator Schmidt rank across every linear bipartition, and an unbounded purification rank (Result \ref{countermulti}). 
We will first show the separation for bipartite states (Sec.~\ref{ssec:bipartite}), and then for multipartite states (Sec.~\ref{ssec:multipartite}).

\subsection{Inequivalence for bipartite classical states}
\lb{ssec:bipartite}

We now show the inequivalence  of the two forms for bipartite classical states, which follows from recent results in convex analysis shown by Gouveia, Parrilo and Thomas \cite{Go12}.  

\begin{result}
\lb{res:ineq}
Let $\rho$ denote a bipartite classical mixed state. Let $D=\trm{OSR}(\rho)$ and  $D'= \trm{rank}_{\trm{puri}}(\rho)$. 
Then $D'$ cannot be upper bounded by a function $f$ that only depends on $D$  (in particular, $f$ does not depend on $\rho$), i.e.
\be
\nexists f: \quad D'\leq f(D)\, .
\ee
\end{result}

\begin{figure}[thb]\centering
\psfrag{A}{(a)}
\psfrag{B}{(b)}
\psfrag{r}{$\rho_{t}$}
\psfrag{P}{$S_{t}$}
\psfrag{i}{$x$}
\psfrag{j}{$y$}
\psfrag{U}{$M$}
\psfrag{V}{$N$}
\psfrag{X}{$X$}
\psfrag{Y}{$Y$}
\psfrag{E}{$E$}
\psfrag{F}{$F$}
\psfrag{e}{$\rk(S_{t})$}
\psfrag{f}{$\rk_{+}(S_{t})$}
\psfrag{g}{$\rk_{\trm{psd}}(S_{t})$}
\psfrag{h}{$\trm{OSR}(\rho_{t})$}
\psfrag{m}{$\rk_{\trm{puri}}(\rho_{t})$}
\psfrag{a}{$A$}
\psfrag{b}{$B$}
\psfrag{c}{$\bar{A}$}
\psfrag{d}{$\bar{B}$}
\includegraphics[width=0.45\textwidth]{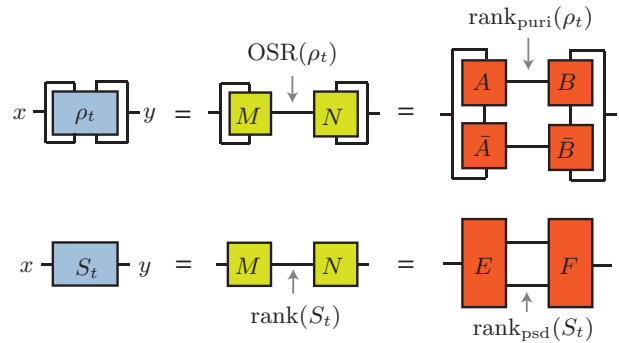}
\caption{
(Left) The classical bipartite state $\rho_{t}$ (equation~\eref{eq:rhot}) (with a joint ket and bra index for the party in state $x$ and the party in state $y$), and below its associated nonnegative matrix $S_{t}$. 
(Middle) 
The operator Schmidt decomposition of $\rho_{t}$ (with dimension $\trm{OSR}(\rho_{t})$) corresponds to the singular value decomposition of $S_{t}$ (with dimension $\trm{rank}(S_{t})$). 
(Right) The local purification form of $\rho_{t}$ (with dimension $\trm{rank}_{\trm{puri}}(\rho_{t})$) corresponds to the positive semidefinite factorization of $S_{t}$ (with dimension $\trm{rank}_{\trm{psd}}(S_{t})$). 
}  
\label{fig:counterexample} 
\end{figure}

Before proving the result, we establish a relation between the MPDO and the local purification form of classical states and two other known decompositions. 
Consider the family of classical bipartite states (i.e.~diagonal in the computational basis $\{|x\ra\}$) of physical dimension $t$ 
\be
\rho_{t} = \sum_{x,y=1}^{t} S_{t}(x,y) |x,y\ra \la x,y|\, ,
\lb{eq:rhot}
\ee
where $S_{t}$ is a nonnegative matrix of size $t\times t$ and $S_{t}(x,y)$ denotes its $x,y$ component. 
Now, the MPDO form of $\rho_{t}$ corresponds to the singular value decomposition of $S_{t}$. Hence, the operator Schmidt rank of $\rho_{t}$ corresponds to the rank of $S_{t}$, $\rk(S_{t})$. 
On the other hand, the local purification form of $\rho_{t}$ corresponds to the positive semidefinite factorization of $S_{t}$, in which $S_{t}$ is expressed as
\be
 \quad S_{t}(x,y)=\tr(E_{x}F_{y})\qquad E_{x},F_{y} \succeq 0\,
\lb{eq:rkpsdP}
\ee
for all $x,y$ (the local purification form is obtained by expressing $E_{x}=A_{x}A_{x}^{\dagger}$ and $F_{y}=B_{y}B_{y}^{\dagger}$). 
Thus, the purification rank of $\rho_{t}$ corresponds to the positive semidefinite rank of $S_{t}$, $\rk_{\trm{psd}}(S_{t})$, which is defined as the minimal $r$ such that there exist matrices 
$E_{x},F_{y}\succeq 0$ of size $r\times r$ such that \eref{eq:rkpsdP} holds (see figure~\ref{fig:counterexample}). 
The positive semidefinite factorization was  very recently introduced in Ref.~\cite{Fi11} and shown to be related to the quantum communication complexity \cite{Fi11} and the quantum correlation complexity of $S_{t}$ \cite{Ja12}. 
In summary, for classical states such as $\rho_{t}$, it holds that 
\footnote{
The nonnegative factorization \cite{Co93} of nonnegative matrices such as $S_{t}$ has also been defined, and has traditionally received more attention, in particular in its connections to classical communication complexity. 
In this case, $S_{t}$ is expressed as a product of two nonnegative matrices $X$ and $Y$. 
The nonnegative rank of $S_{t}$, $\rk_{+}(S_{t})$, is the minimal $r$ such that there exist 
$X,Y\geq 0$ of size $t\times r$ and  $r\times t$, respectively, such that $S_{t}=XY$.
Note that this factorization cannot be defined for general quantum states.
}
\be
\ba{l}
\trm{OSR}(\rho_{t})= \rk(S_{t})\vspace{1mm}\\
\rk_{\trm{puri}}(\rho_{t})=\rk_{\trm{psd}}(S_{t})\, .
\ea
\lb{eq:qranks}
\ee

\begin{figure}[thb]\centering
\psfrag{S}{$S_{6}=$}
\psfrag{0}{0}
\psfrag{1}{1}
\psfrag{2}{2}
\includegraphics[width=0.45\textwidth]{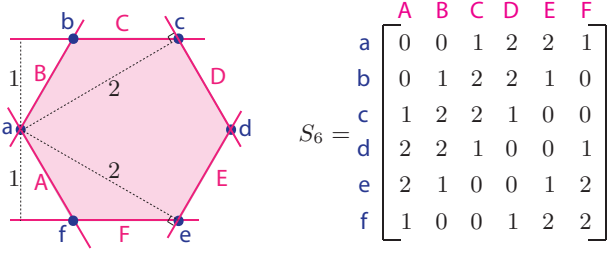}
\caption{
The slack matrix of the regular hexagon $S_{6}$ contains in its $(i,j)$ entry the distance from vertex $i$ (here labeled with lowercase letters) to hyperplane $j$ (labeled with uppercase letters). Note that the matrix is circulant.  
}  
\label{fig:6gon} 
\end{figure}

Now we are ready to prove Result \ref{res:ineq}.
\begin{proof}
We consider classical bipartite states of the form \eref{eq:rhot}, and we focus  on a class of nonnegative matrices $S_{t}$ called slack matrices of polytopes, defined as follows (see e.g.~\cite{Go12}). 
A convex polytope is defined as the intersection of a finite set of halfspaces $\{h_{j}(x)\leq b_{j}\}$, or as the convex hull of a set of vertices $\{v_{i}\}$ \cite{Zi07}. 
Its slack matrix $S$ is defined so that its $(i,j)$ entry contains the distance from hyperplane $j$ to vertex $i$, i.e.~$S(i,j)=b_{j}-h_{j}(v_{i})$ (see figure~\ref{fig:6gon}). 

Now we let $S_{t}$ be the slack matrix of the two-dimensional regular polytope with $t$ vertices (and thus also $t$ faces), called the regular $t$-gon
\footnote{If centered at the 0 point, the regular $t$-gon is defined by 
$P_{t}=\textrm{conv}\{(\cos(2\pi k/t),\sin(2\pi k/t))
:  0\leq k <t\} \subseteq \mathbb{R}^{2}$ \cite{Zi07}.}. 
Gouveia, Parrilo and Thomas show that \cite{Go12} \footnote{More precisely, they show that
$\Omega(\log t)\leq \rk_{\mathrm{psd}}(S_{t})\leq \log t$. 
We use the $O$ and $\Omega$ notation as customary, see e.g.~\cite{Mo11}.}
% $\rk (S_{t})=2+1$, since the polytope lives in 2 dimensions \cite{Go12b} 
\be
\ba{l}
\rk (S_{t})=3\qquad \forall t  \vspace{1mm}\\ 
\rk_{\mathrm{psd}}(S_{t})\sim \log t\, .
\ea
\lb{eq:cc}
\ee
Using the equivalences \eref{eq:qranks}, this implies that
\begin{equation}
\ba{l}
\trm{OSR}(\rho_{t})=3 \qquad \forall t\vspace{1mm}\\
\rk_{\trm{puri}}(\rho_{t})\sim \log t \, .
\ea
\lb{eq:qc}
\end{equation}
That is, the operator Schmidt rank of $\rho_{t}$ is constant for all $t$, whereas its purification rank grows unboundedly with $t$. 
It follows that there does not exist an upper bound of $\rk_{\mathrm{puri}}(\rho_{t})$ that depends only on $\trm{OSR}(\rho_{t})$.
\end{proof}

\subsection{Inequivalence for multipartite classical states}
\lb{ssec:multipartite}

Now we  show a more general form of separation, 
namely we provide classical multipartite states with a constant operator Schmidt rank across every linear bipartition, and an unbounded purification rank.

\begin{result}
\lb{countermulti}
Let $\rho$ denote a mutipartite classical mixed state. Let $D=\trm{OSR}(\rho)$ and  $D'=\trm{rank}_{\trm{puri}}(\rho)$. 
Then $D'$ cannot be upper bounded by a function $f$ that depends only of $D$ (in particular, $f$ is independent of $\rho$), i.e.
\be
\nexists f: \quad D'\leq f(D)\, .
\ee
\end{result}

\begin{proof}
Consider the family of states $\rho_{t}$ (equation~\eref{eq:rhot}) with $t=2^{m}$ (with natural $m$).
That is, $S_{t}$ is the slack matrix of the square, octagon, etc. 
Let the binary representations of the row index $x$ and column index $y$ be $(x_{1},\ldots,x_{m})$ and  $(y_{1},\ldots,y_{m})$, respectively.

Since the polygon is regular, $S_{t}$ is a circulant matrix. 
% (that is, $S(x,y)=S(x-y)$)
Hence, it is diagonalized by the Fourier transform $F_{t}$, with components $F_{t}(x,y)=\omega^{xy}$, where $\omega=\exp(i2\pi/2^{m})$. That is,  
\be
S_{t}(x,y)=
\sum_{\alpha=1}^{3}
L^{\alpha}_{x_{1},\ldots,x_{m}} 
R^{\alpha}_{y_{1},\ldots,y_{m}}\, ,
\lb{eq:S_{t}}
\ee
where 
\be
\ba{l}
L^{\alpha}_{x_{1},\ldots,x_{m}} = \exp\left[-2\pi i \alpha \: 0.x_{1}\ldots x_{m} \right]\lambda_{\alpha}  \vspace{1.5mm}\\
R^{\alpha}_{y_{1},\ldots,y_{m}}=\exp\left[2\pi i \alpha \: 0.y_{1}\ldots y_{m} \right]\, , 
\ea
\ee
and $\{\lambda_{\alpha}\}$ are the eigenvalues of $S_{t}$. 
Now we decompose the tensors $L$ and $R$ into smaller tensors, 
\be
\ba{l}
L^{\alpha}_{x_{1}\ldots x_{m}}=
(M_{1})_{x_{1}}^{\alpha} 
\ldots (M_m)_{x_{m}}^{\alpha,\alpha} \vspace{1.5mm}\\
R^{\alpha}_{y_{1}\ldots y_{m}}=
(M_{m+1})_{y_{1}}^{\alpha,\alpha} 
\ldots (M_{2m})_{y_{m}}^{\alpha} \, ,
\ea
\ee
where each $M_{k}$ depends only on one bit ($x_{k}$ or $y_{k}$),
\be
\ba{l}
(M_k)^{x_{k}}_{\alpha,\alpha}=\exp\left[-2\pi i\alpha\: \frac{x_{k}}{2^{k}} \right]
 \qquad  1< k < m\vspace{1.5mm}\\
(M_m)^{x_{m}}_{\alpha,\alpha}=\exp\left[-2\pi i\alpha\: \frac{x_{m}}{2^{m}} \right]\lambda_{\alpha} \vspace{1.5mm}\\
(M_k)^{y_{k-m}}_{\alpha,\alpha}=\exp\left[2\pi i \alpha\: \frac{y_{k-m}}{2^{k-m}} \right]
\qquad m<k < 2m\, .
\ea
\ee
That is, each $M_{k}$ is a $3\times 3$ diagonal matrix, except for $M_{1}$ and $M_{2m}$, which are a row and a column vector, respectively, defined analogously.
This shows that $\rho_{t}$ has operator Schmidt rank 3 across every linear bipartition.

\begin{figure}[thb]\centering
\psfrag{1}{$x_{1}$}
\psfrag{2}{$x_{2}$}
\psfrag{3}{$y_{1}$}
\psfrag{4}{$y_{2}$}
\psfrag{p}{$\rho_{t}$}
\psfrag{R}{$R$}
\psfrag{L}{$L$}
\psfrag{a}{$M_{1}$}
\psfrag{b}{$M_{2}$}
\psfrag{c}{$M_{3}$}
\psfrag{d}{$M_{4}$}
\psfrag{=}{$=$}
\psfrag{t}{3}
\psfrag{A}{$B$}
\psfrag{C}{$\bar{B}$}
\psfrag{B}{$C$}
\psfrag{D}{$\bar{C}$}
\psfrag{x}{$A_{1}$}
\psfrag{y}{$A_{2}$}
\psfrag{z}{$A_{3}$}
\psfrag{w}{$A_{4}$}
\psfrag{X}{$\bar{A_{1}}$}
\psfrag{Y}{$\bar{A_{2}}$}
\psfrag{Z}{$\bar{A_{3}}$}
\psfrag{W}{$\bar{A_{4}}$}
\psfrag{l}{$\log t$}
\includegraphics[width=0.49\textwidth]{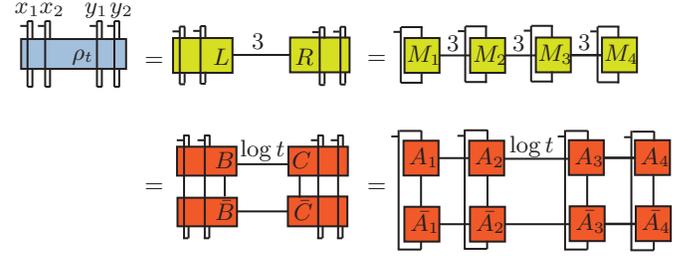}
\caption{The classical state $\rho_{t}$ (equation~\eref{eq:rhot}), where $S_{t}$ is the slack matrix of the regular $t$-gon, shows the separation of the MPDO form and the local purification form. 
(Middle) In the bipartite case, $\rho_{t}$ has  operator Schmidt rank 3 for all $t$, and purification rank $\sim \log t$. 
(Right) In the multipartite case, for $t=2^{m}$, $\rho_{t}$ has  operator Schmidt rank 3 across every linear bipartition, and purification rank $\sim \log t$ at least across one bipartition.
}  
\label{fig:ft} 
\end{figure}

We know from equation~\eref{eq:qc} that the purification rank of $\rho_{t}$ along the $x$ vs.~$y$ bipartition grows unboundedly like $\log t$. 
Clearly, a small purification rank across any bipartition would imply a small purification rank across the $x$ vs.~$y$ bipartition. 
Thus, the purification rank is unbounded at least across one bipartition (see figure~\ref{fig:ft}).
\end{proof}

This shows that a small operator Schmidt rank across all linear bipartitions does not imply a small purification rank.
%Note that the middle cut (the one with $x$ vs.~$y$) typically has the largest purification rank. 

Let us make a final remark. Note that a possible purification of $\rho_{t}$ is 
\be
&&\rho_{t}=\textrm{tr}_{\trm{anc}}|\varphi_{t}\ra \la \varphi_{t}| \nn\\
&&|\varphi_{t}\ra = \sum_{x,y=1}^{t} \sqrt{S_{t}(x,y)}|x, x,y,y\ra\, ,
\lb{eq:psitilde}
\ee
where the second and fourth index refer to the ancillary states. 
Thus,  $\rk_{\textrm{puri}}(\rho_{t})\leq \trm{SR}(\varphi_{t})$. Above we have seen that $\rk_{\textrm{puri}}(\rho_{t})\sim \log t$, thus the Schmidt rank of $|\varphi_{t}\ra$ grows with $t$, and so does its preparation cost \footnote{In the sense of the dimension of the ancilla required to prepare the state in a sequential scheme \cite{Sc05c}.}. 
Now consider the state obtained by taking the square of the coefficients of $|\varphi_{t}\ra$, 
\be
|\varphi_{t}^{\odot 2}\ra := \sum_{x,y=1}^{t}S_{t}(x,y)|x,x,y,y\ra 
\lb{eq:psi}
\ee
From Result \ref{countermulti} it follows that the Schmidt rank of this state across any linear bipartition is 3. Hence its preparation cost is constant with $t$ \cite{Sc05c}. 
Thus, we see that transforming \eref{eq:psi} to \eref{eq:psitilde}, i.e.~taking the Hadamard square root of the coefficients, may have a high cost in the Schmidt rank, and thus in the preparation cost of a pure state.

\section{Purification methods}
\lb{sec:purif}

We will now present two constructive purification methods: the sos polynomial method (Sec.~\ref{ssec:1method}) and the eigenbasis method (Sec.~\ref{ssec:2method}). 
Both are applicable to all multipartite mixed states, and can be used to construct exact and approximate purifications.
We will compare both approximation methods and see that they are complementary for various eigenvalue distributions in Sec.~\ref{ssec:compare}.

\subsection{Sos polynomial method}
\lb{ssec:1method}

We will first present the idea of the sos polynomial method (Sec.~\ref{sssec:1idea}),
and then explain how to use it to construct exact (Sec.~\ref{sssec:1exact}) and approximate purifications  (Sec.~\ref{sssec:1approx}).

\subsubsection{The idea}
\lb{sssec:1idea}

The idea of the sos polynomial method is the following: given a mixed state $\rho$, we construct a purifying state as a sum of powers of $\rho$ (up to certain degree), where each power is attached to an ancillary state. If the degree is large enough, there exists a choice of the ancillary states such that this purifying state is an exact purification for $\rho$. 
If the degree is not large enough, 
one can find the ancillary states with an ansatz of sos polynomials or with an SDP.

\begin{figure}[thb]\centering
\psfrag{P}{$|\Psi_{k}\ra=$}
\psfrag{0}{$|\rho^{0}\rangle$}
\psfrag{1}{$|\rho^{1}\rangle$}
\psfrag{k}{$|\rho^{k-1}\rangle$}
\psfrag{a}{$a_{0}$}
\psfrag{b}{$a_{1}$}
\psfrag{c}{$a_{2}$}
\psfrag{d}{$a_{k-1}$}
\psfrag{A}{$\bar a_{0}$}
\psfrag{B}{$\bar a_{1}$}
\psfrag{C}{$\bar a_{2}$}
\psfrag{D}{$\bar a_{k-1}$}
\psfrag{o}{$\rho^{0}$}
\psfrag{p}{$\rho^{1}$}
\psfrag{q}{$\rho^{2}$}
\psfrag{z}{$\rho^{2(k-1)}$}
\psfrag{+}{+}
\psfrag{...}{$\ldots$}
\psfrag{r}{$\sigma_{k}=$}
\includegraphics[width=0.49\textwidth]{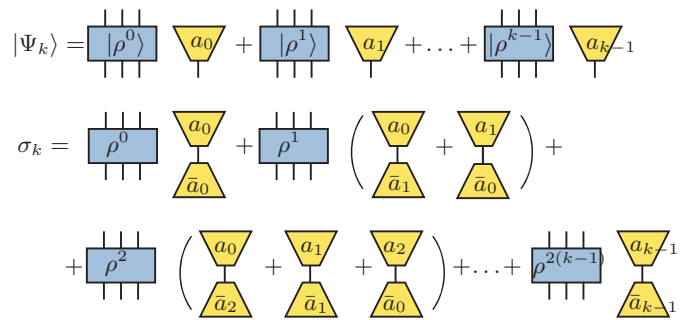}
\caption{The sos polynomial method constructs a state $|\Psi_{k}\ra$ as a sum of powers of $\rho$, from 0 to $k-1$, each attached to an ancillary state, $|a_{l}\ra$. 
$|\Psi_{k}\ra$ is used as purification of $\sigma_{k}$, which has the same eigenvectors as $\rho$, but its eigenvalues are a sos polynomial of the eigenvalues of $\rho$.
If $k=m$, where $m$ is the number of different nonnegative eigenvalues of $\rho$, $\sigma_{k}$ can be  an exact purification of $\rho$. 
If $k<m$, one can search for the $\sigma_{k}$ closest to $\rho$ with an SDP.
}  
\label{fig:ideamethod1} 
\end{figure}

Specifically, consider a multipartite density matrix $\rho$ of size $d^{N} \times d^{N}$, with spectral decomposition
\be
\rho=\sum_{i=1}^{n}\lambda_{i}|\phi_{i}\ra\la\phi_{i}| + 
\sum_{i=n+1}^{d^{N}} \lambda_{i}|\phi_{i}\ra\la \phi_{i}|\, , 
\lb{eq:eig}
\ee
where $\lambda_{i}=0$ for $i>n$. 
We construct a purifying state $|\Psi_{k}\ra$ as a sum of powers of $\rho$, from 0 to $k-1$, where each power is attached to an ancillary state (see figure~\ref{fig:ideamethod1}), 
\be 
|\Psi_{k}\ra=\sum_{l=0}^{k-1}|\rho^{l}\ra_{{\scriptsize KB}}\otimes |a_{l}\ra_{{\scriptsize A}}\, ,
\lb{eq:Psi}
\ee
where $|\rho\ra$ denotes a vectorized matrix $\rho$ (i.e.~given 
$\rho=\sum_{i,j}\rho_{i,j}|i\ra\la j|$ we define
$|\rho\ra_{KB}=\sum_{i,j}\rho_{i,j}|i\ra_{K}|j\ra_{B}$). 
We use $|\Psi_{k}\ra$ as the purifying state of a density matrix $\sigma_{k}$, 
\be
\sigma_{k}&=&\tr_{BA}|\Psi_{k}\ra\la\Psi_{k}| 
=\sum_{i=1}^{d^{N}} 
p_{k}(\lambda_{i})
\: |\phi_{i}\ra\la\phi_{i}|\, .
\lb{eq:sigma}
\ee
Here $p_{k}$ is a polynomial that can be written as 
\be
p_{k}(\lambda)=(1, \lambda, \ldots, \lambda^{k-1})\: R_{k}\: (1, \lambda, \ldots, \lambda^{k-1})^{T}
\, ,
\lb{eq:pk}
\ee
where $R_{k}$ is a positive semidefinite matrix which is the Gram matrix of the ancillary states, i.e.~its $(i,j)$ component is $R_{k}(i,j)=\la a_{i}|a_{j}\ra$. 
%$R_{k}$ is called the Gram matrix of the sos representation
This polynomial is \emph{sum of squares} (sos), as can be readily seen by writing $R_{k}=A^{T}A$,
\be
&&p_{k}(\lambda) = \sum_{u=1}^{r}y_{u}(\lambda)^{2}\\
&&y_{u}(\lambda) = A_{u} (1, \lambda, \ldots, \lambda_{k-1})^{T}\, ,
\ee
where $A_{u}$ denotes the $u$th row of $A$, and $r=\textrm{rank}(R_{k})$. % (thus $r\leq k$).
Note that, since the polynomial is univariate, the set of sos polynomials is identical  to the set of nonnegative polynomials (i.e.~with the property $p(x)\geq 0$ for all $x$) \cite{Bl13}. 

%Note that, in the range of $\rho$, $\sigma_{k}$ has the same eigenvectors as $\rho$, but its eigenvalues are $p_{k}(\lambda_{i})$ (equation~\eref{eq:sigma}), and the support of $\sigma_{k}$ extends to the entire physical dimension $d^{N}$ (due to the zeroth order term in \eref{eq:Psi}).

Now, by construction, the purification rank of $\sigma_{k}$ is at most the Schmidt rank of $|\Psi_{k}\ra$ (equation~\eref{eq:sigma}). Denoting $\trm{OSR}(\rho)=D$, this is at most $1+D+D^{2}+\ldots+D^{k-1}$ (equation~\eref{eq:Psi}), and thus
\be
\rk_{\trm{puri}}(\sigma_{k})\leq \textrm{SR} (\Psi_{k}) \leq \frac{D^{k}-1}{D-1}\, .
\lb{eq:rpsigma}
\ee

We are interested in making $\sigma_{k}$  as close as possible to $\rho$. 
We will show that if $k=m$, where $m$ is the number of different nonnegative eigenvalues of $\rho$ 
\footnote{By nonnegative eigenvalues, we mean that the 0 should also be counted as an eigenvalue. That is, if $\rho$ is rank deficient, $m$ is the number of different non-zero eigenvalues plus one.}, 
there exists ancillary states such that $p_{k}(\lambda_{i})=\lambda_{i}$ for all $i$, and thus $\sigma_{m}=\rho$, i.e.~$\sigma_{m}$ is an exact purification of  $\rho$ (Sec.~\ref{sssec:1exact}). 
If, on the contrary, $k<m$, one 
can choose $k-1$ points and construct the $p_{k}$ that passes through them, or one can find the $p_{k}$ that minimizes the distance between $\sigma_{k}$ and $\rho$  (in trace norm) with an SDP (Sec.~\ref{sssec:1approx}).

\subsubsection{Exact case}
\lb{sssec:1exact}

We now show how to build exact purifications with the sos polynomial method. 

\begin{result}
Let $\rho$ denote a multipartite density matrix with $m$ different nonnegative eigenvalues. 
Let $D=\textrm{OSR}(\rho)$ and $D'=\rk_{\textrm{puri}}(\rho)$.
The sos polynomial method provides an exact purification of $\rho$ with
\be
D' \leq \frac{D^{m}-1}{D-1}\, .
\ee
\label{lem:1}
\end{result}

\begin{proof}
To ease the notation we consider that the different nonnegative eigenvalues are the first $m$ 
(i.e.~$\lambda_{i}\neq \lambda_{j}$ for $i\neq j$ and $i,j=1,\ldots,m$). 
Consider the construction of the purifying state $\sigma_{k}$ as explained in Sec.~\ref{sssec:1idea}. 
Define the vector $|v_{k}^{i}\ra$ as
\be
|v_{k}^{i}\ra = (1, \lambda_{i}, \ldots, \lambda_{i}^{k-1})^{T}\, , 
\ee
so that the sos polynomial evaluated at $\lambda_{i}$ can be written as $p_{k}(\lambda_{i})= \la v_{k}^{i}|R_{k}|v_{k}^{i}\ra$ (compare with equation~\eref{eq:pk}). 
Now we choose $k=m$. 
The important observation is that  the set of vectors $\{|v^{i}_{m}\ra\}_{i=1}^{m}$ is linearly independent (since only different eigenvalues are considered).
Hence, there exists another set $\{ |w_{m}^{j}\ra\}_{j=1}^{m}$ which is biorthogonal to it, i.e.~$\la v_{m}^{i}|w_{m}^{j}\ra=\delta_{ij}$ for all $i,j$.
Then we choose $R_{m}$ as follows:
\be
R_{m}=\sum_{j=1}^{m}\lambda_{j}|w_{m}^{j}\ra \la w_{m}^{j}|.
\lb{eq:R}
\ee
This satisfies that $p_{m}(\lambda_{i})=\la v_{m}^{i}|R_{m}|v_{m}^{i}\ra= \lambda_{i}$ for all $i$. From equation~\eref{eq:sigma} it follows that $\sigma_{m} =\rho$ and thus $\sigma_{m}$ is an exact purification of $\rho$. 
Finally, using equation~\eref{eq:rpsigma} with $k=m$, the claim of the result follows.
 \end{proof}

Note that Result \ref{lem:1} depends on the number of different nonnegative eigenvalues $m$ because $\sigma_{k}=\rho$ requires that $p(\lambda_{i})=\lambda_{i}$ for all $i$, and these are only $m$ independent conditions.

\subsubsection{Approximate case: SDP} 
\lb{sssec:1approx}

Building approximate purifications with the sos polynomial method can be done analytically or numerically, as we show next.

\begin{result}
\lb{lem:1approx}
Let $\rho$ denote a multipartite density matrix 
of  dimension $d^{N}\times d^{N}$, with $\rank(\rho)=n$.  
Let $D=\textrm{OSR}(\rho)$. 
The sos polynomial method provides a density matrix $\sigma_{k}$ with
\be 
\trm{rank}_{\trm{puri}}(\sigma_{k})\leq \frac{D^{k}-1}{D-1}
\lb{eq:1approx}
\ee
and
\be
||\rho-\sigma_k||_{1} = \sum_{i=1}^{n}|\lambda_{i}-p_{k}(\lambda_{i})|+ 
(d^{N}-n)p_{k}(0)
\, , 
\lb{eq:1approxdist}
\ee
where $p_{k}$ is a sos polynomial of degree $2(k-1)$.
$p_{k}$ can be constructed by choosing $k-1$ nonnegative points through which it must pass. 
Alternatively, the optimal $p_{k}$ can be found with an SDP.
\end{result}

%A sos polynomial $p_{k}$ of degree $2(k-1)$ has at most $k-1$ real roots (since roots of sos polynomials have even multiplicity). 

\begin{proof}

\emph{Sos polynomial that passes through $k-1$ points}. 
We construct the sos polynomial $p_{k}$ by letting it pass through  $k-1$ chosen points $\{\mu_{1},\ldots,\mu_{k-1}\}$. We use the Lagrange basis ``squared'' to this end:
\be
\lb{eq:lagrange}
p_{k}(\lambda) &=& \sum_{j=1}^{k-1} \mu_{j} l_{j}(\lambda)\\
 l_{j}(\lambda)&= &
\prod_{i=1, i\neq j}^{k-1}
\left(\frac{\lambda - \mu_{i}}{\mu_{j}-\mu_i}\right)^{2}\, , 
\ee
where we have omitted the dependence of $l_{j}$ on $k$. 
This satisfies $p_{k}(\mu_{j})=\mu_{j}$ for $j=1,\ldots,k-1$.
Note that the degree of $p_{k}$ is $2(k-1)$ as required. 
The distance \eref{eq:1approxdist} depends on the points $\{\mu_{i}\}$, thus the difficulty lies in choosing them. 

\emph{The optimization problem as an SDP.} 
Let $\{\lambda_{i}\}_{i=1}^{d^{N}}$ denote the eigenvalues of $\rho$, with $\lambda_{i}=0$ for $n<i\leq d^{N}$.
We search for the sos polynomial $p_{k}$ that minimizes the distance \eref{eq:1approxdist}, i.e.~for the 
positive semidefinite matrix $R_{k}$ that minimizes it (see equation~\eref{eq:pk}).
The optimization problem thus reads 
\be\lb{eq:sdp}
&&\min\: \sum_{i=1}^{d^{N}} |\lambda_{i}-\la v_k^i|R_{k}|v_k^i\ra|\\
&&\trm{s.t. } \: R_{k}\succeq 0\, .\nn
\ee
The objective function can be made linear by introducing the slack variables $z$, 
\be\lb{eq:sdp2}
&&\min \: \sum_{i=1}^{d^{N}} z_{i}\\
&&\trm{s.t. } \:  z_{i} \geq  \lambda_{i}-\la v_k^i|R_{k}|v_k^i\ra\quad \qquad i=1,\ldots,d^{N} \nn\\
&&\qquad  z_{i} \geq  -\lambda_{i}+\la v_k^i|R_{k}|v_k^i\ra  \qquad \: i=1,\ldots,d^{N} \nn\\
&&\qquad R_{k}\succeq 0\, ,\nn
\ee
The optimization variables are now $R_{k}\succeq0$ and $z\geq 0$, and the constraints are linear in them.
Thus, this is an SDP optimization problem (see Appendix \ref{app:SDP} for the precise formulation).
\end{proof}

In words, the SDP searches for the sos polynomial $p_{k}$ (of degree $2(k-1)$) whose distance to the eigenvalue distribution is minimal in trace norm.  
This formulation is consistent with the fact that optimization over sos polynomials can be done with SDPs \cite{Bl13}.

Note that the non-trivial condition is that the polynomial be sos (equivalently, nonnegative), since otherwise one could take $p(\lambda)=\lambda$.
Approximating  $\lambda$ for $0\leq \lambda\leq \lambda_{\trm{max}}$ and another nonnegative function elsewhere has the problem that the function at $\lambda=0$  is non analytical.

We remark that there may exist exact solutions for $m/2 < k < m$, since $p_{k}(\lambda)-\lambda$ has degree $2(k-1)$, and thus can have $2(k-1)$ real roots. 
However, we only know how to construct the exact solution for the case $k=m$ (with the idea of the proof of Result \ref{lem:1}).

\begin{remark}
\emph{Non-orthogonal ancillary states}. 
Observe that the exact solution for $k=m$ involves non-orthgonal ancillary states, since $R_{m}$ is non-diagonal (equation~\eref{eq:R}). 
This is so because the basis $\{|v_{m}^{i}\ra\}_{i=1}^{m}$ is not orthogonal, and neither is the biorthogonal basis $\{|\omega_{m}^{j}\ra\}_{j=1}^{m}$.
More generally, the solution for $k<m$ also involves non-orthogonal ancillary states, 
since orthogonal states result in a diagonal $R_{k}$, which renders a polynomial only even powers with non-negative coefficients (i.e.~a monotonously increasing polynomial for positive $\lambda$). 
In contrast, a non-diagonal $R$ yields polynomials with possibly odd powers with negative coefficients, thus with various minima, rendering a better approximation of $\lambda$ in the desired interval. 
\end{remark}

\begin{remark}
\lb{rem:realanc}
\emph{Real ancillary vectors}.
The ancillary vectors $\{|a_{j}\ra\}$ can be taken real without loss of generality. 
To see this, first write the polynomial (equation~\eref{eq:pk})  as 
\be
p_{k}(\lambda)
= \sum_{s,t=0}^{k-1}R_{k}(s,t)\lambda^{s+t}
= \sum_{l=0}^{2(k-1)} c_{l}
\lambda^{l}
\ee
where $c_{l} = \sum_{s+t=l}R_{k}(s,t)$.   
That is, each coefficient $c_{l}$ is the sum of the $l$th antidiagonal of $R_{k}$. Thus $c_{l}$ depends on diagonal elements of $R_{k}$, which are non-negative, or on the sum of an element and its transposed ($R_{k}(s,t)+R_{k}(t,s)$), which is real. Therefore $c_{l}$ only depends on the real part of $R$. 
If $R\succeq 0$, then $\textrm{Re}(R)$ is positive semidefinite for all real vectors $|r\rangle$, since 
$\la r | \textrm{Re}(R) + i \textrm{Im}(R)| r \ra \geq 0$, which implies $i\la r |  \textrm{Im}(R)| r \ra =0$.
Since we only consider contractions with real vectors (namely $|v_{k}^{i}\ra$), we can restrict $R$ to its real part. Thus, its spectral decomposition reads $R=ODO^{T}$, where $O$ is orthogonal, and $D$ is diagonal and nonnegative. 
This readily yields $R=O\sqrt{D}\sqrt{D}O^{T}=AA^{T}$, where the $i$th row of $A$ contains the coefficients of the ancillary state $\la a_{i}|$ expressed in the eigenbasis of $R$, which are real.
\end{remark}

%This generally works well if the eigenvalues cluster around a few points, like in rapidly decaying distributions. 
%However, in general, it is hard to choose the $k$ points through which the polynomial should pass. For this case, we have formulated the search for the best sos polynomial of degree $2(k-1)$ as an SDP optimization problem (Sec.~\ref{sssec:1approx}).

While the exact sos polynomial method depends on $m$ (Result \ref{lem:1}) and the approximate on $d^{N}$ (Result \ref{lem:1approx}), we will see in Sec.~\ref{ssec:compare} that in practice there exists good ans\"atze of sos polynomials which make it independent of both.
We will also compare   this approximate method with the eigenbasis method, which we present next.

%===============================
%===============================
\subsection{Eigenbasis method}
\lb{ssec:2method}
%===============================
%===============================

Now we turn to the eigenbasis method, for which we present the main idea (Sec.~\ref{sssec:2idea}), and how to use it to construct exact (Sec.~\ref{sssec:2exact}) and approximate purifications (Sec.~\ref{sssec:2approx}). 
%which is based on bounding the Schmidt rank of the eigenbasis of the density matrix 
%Its exact version yields a purification rank that scales multiplicatively with the number of eigenvalues  % and its approximate version works by discarding the smallest eigenvalues . 

%===============================
\subsubsection{The idea}
\lb{sssec:2idea}

The idea of the eigenbasis method is to consider the standard purification of $\rho$, which is the spectral decomposition, and upper bound the Schmidt rank of each eigenstate of $\rho$. 
This is done by constructing a basis of the range of $\rho$ where each basis element is the image (under the map $\rho$) of a product state. Thus, each has Schmidt rank at most $D$, where $D$ is the operator Schmidt rank of $\rho$. 
Expressing each eigenstate in terms of this basis, we see that each can have Schmidt rank at most $Dn$, where $n$ is the number of non-zero eigenvalues of $\rho$.
Thus the purification rank of $\rho$ is at most $Dn^{2}$. 

\begin{figure}[thb]\centering
\psfrag{r}{$\rho$}
\psfrag{s}{$\rho(|\mu_{\gamma}\ra,|\nu_{\gamma}\ra)=$}
\psfrag{x}{$\chi$}
\psfrag{P}{$P$}
\psfrag{Q}{$Q$}
\psfrag{R}{$R$}
\psfrag{p}{$\phi$}
\psfrag{q}{$\bar{\phi}$}
\psfrag{D}{$D$}
\psfrag{n}{$n$}
\psfrag{b}{$p_{1}$}
\psfrag{c}{$p_{2}$}
\psfrag{d}{$p_{3}$}
\psfrag{e}{$g$}
\psfrag{=}{$=$}
\psfrag{a}{$\alpha$}
\psfrag{i}{$i$}
\psfrag{S}{$\textrm{SR}$}
\psfrag{l}{$\leq D$}
\psfrag{A}{(a)}
\psfrag{B}{(b)}
\psfrag{C}{(c)}
\includegraphics[width=0.49\textwidth]{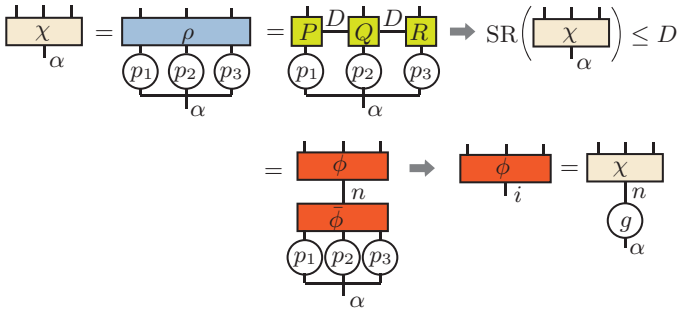}
\caption{The eigenbasis method constructs a state $|\chi_{\alpha}\ra$ as the image (under $\rho$) of the product state $\otimes_{i=1}^{N}|p_{i}\ra_{\alpha}$, here shown for $N=3$.
(Upper line) 
Using the operator Schmidt decomposition of $\rho$ we see that $\textrm{SR}(\chi_{\alpha})\leq D$. 
(Lower line)
 Using the spectral decomposition of $\rho$, we express each eigenstate $|\phi_{i}\ra$ as a linear combination of (at most $n$) $|\chi_{\alpha}\ra$  (equation~\eref{eq:g}), and upper bound the Schmidt rank of each eigenstate by $Dn$. 
}  
\label{fig:ideamethod2} 
\end{figure}

To be more precise, consider the standard purification of $\rho$, obtained from its spectral decomposition (see equation~\eref{eq:eig}) as 
\be
&&\rho = \tr_{A}|\Psi\ra\la \Psi| \\
&&|\Psi\ra = \sum_{i=1}^{n}\sqrt\lambda_{i}|\phi_{i}\ra_{S}|i\ra_{A}\, .
\lb{eq:2purif}
\ee
By construction, the purification rank of $\rho$ is at most the Schmidt rank of $|\Psi\ra$, which is at most $n$ times the maximum Schmidt rank of $|\phi_{i}\ra$, 
\be
\textrm{rank}_{\textrm{puri}}(\rho) \leq \textrm{SR}(\Psi)
\leq n \max_{i} \textrm{SR}(\phi_{i})\, .
\lb{eq:2rp}
\ee
Our goal is to upper bound $\max_{i}\textrm{SR}(\phi_{i})$ as a function of $D$.
To this end, we build a basis of the range of $\rho$ where each basis element $|\chi_{\alpha}\ra$ is the image (under the map $\rho$) of a certain product state $|p_{\alpha}\ra$ (see figure~\ref{fig:ideamethod2}),
\be
&&|\chi_{\alpha}\ra=\rho \: |p_{\alpha}\ra \\
&&|p_{\alpha}\ra=\bigotimes_{i=1}^{N} |p_{\alpha_{i}}\ra_{i}\, , 
\lb{eq:proj}
\ee
where $\alpha=(\alpha_{1},\ldots,\alpha_{N})$, $\alpha_{i}$ is a label of the $i$th product state, and the subindex outside the ket denotes the subsystem that the state is describing. 

Now, consider the MPDO form of $\rho$ as in equation~\eref{eq:osd}, with operator Schmidt rank $D$. 
The Schmidt rank of $|p_{\alpha}\ra$ is one, 
and applying $\rho$ to $|p_{\alpha}\ra$ increases the Schmidt rank by at most $D$. 
It follows that 
\be
\textrm{SR} (\chi_{\alpha}) \leq D\, .
\ee

On the other hand, we consider the spectral decomposition of $\rho$ (equation~\eref{eq:eig}), and use the eigenstates as a basis of the range. 
In particular, we express $|\chi_{\alpha}\ra$ in terms of this basis, with coefficients $f_{i\alpha}$ of the linear combination, 
\be
|\chi_{\alpha}\ra&=&\sum_{i=1}^{n} |\phi_{i}\ra 
\underbrace{\lambda_{i}\la \phi_{i} | p_{\alpha}\ra }_{f_{i\alpha}}\, .
\lb{eq:chi}
\ee
Since the range has dimension $n$, there are at most $n$ linearly independent $|\chi_{\alpha}\ra$.  
The idea of the exact result (Sec.~\ref{sssec:2exact}) is to show that one can invert the relation \eref{eq:chi} and express $|\phi_{i}\ra$ as a linear combination of (at most $n$)  $|\chi_{\alpha}\ra$. 
The approximate method discards the smallest eigenvalues, and applies the exact result to the truncated density matrix (Sec.~\ref{sssec:2approx}).

%===============================
\subsubsection{Exact case}
\lb{sssec:2exact}

We now show how to use the eigenbasis method  to construct an exact purification. 

\begin{result}
\lb{lem:2exact}
Let $\rho$ denote a multipartite density matrix with $\trm{rank}(\rho)=n$. 
Let $D=\textrm{OSR}(\rho)$ and $D'=\rk_{\textrm{puri}}(\rho)$. 
Then the eigenbasis method constructs an exact purification of $\rho$ with 
\be
D' \leq D  n^{2}\, .
\ee
\label{lem:2}
\end{result}

\begin{proof} 	
Let $\rho$ be a multipartite density matrix with  spectral decomposition as in \eref{eq:eig} (thus with $\trm{rank}(\rho)=n$). 
Building upon Sec.~\ref{sssec:2idea}, we only need to show that there exist $n$ product states $\{|p_{\alpha}\ra\}_{\alpha=1}^{n}$ such that their images under $\rho$, $\{|\chi_{\alpha}\ra=\rho |p_{\alpha}\ra\}_{\alpha=1}^{n}$,
 form a basis of the range of $\rho$. 
Take a product basis  $\{|x\ra\}$ with $x=1,\ldots, d^N$, which spans the whole space.
Then $\{\rho |x\ra\}$ spans the range of $\rho$. Then, we can select a basis $\rho|x_s\ra$ of the range of $\rho$ with $s=1,\ldots, n$.
We call the latter product states $|p_\alpha\ra$ with $\alpha=1,\ldots,n$.

This implies that the coefficient matrix $(f_{i\alpha})$ (equation~\eref{eq:chi}) is full-rank and hence can be inverted. 
We denote the elements of the inverse by $g_{\alpha j}$, i.e.~$\sum_{\alpha=1}^{n}f_{i\alpha}g_{\alpha j}=\delta_{ij}$, and we invert equation~\eref{eq:chi} to express the eigenvector $|\phi_{i}\ra$ as a linear combination of the vectors $|\chi_{\alpha}\ra$,  
\be
|\phi_{i}\ra=\sum_{\alpha=1}^{n} g_{\alpha i} |\chi_{\alpha}\ra\, .
\lb{eq:g}
\ee
Since $\trm{SR} (\chi_{\alpha})\leq D$, we obtain
\be
\trm{SR}  (\phi_{i}) \leq Dn \, ,
\ee for all $i$.
Finally, using \eref{eq:2rp}, the claim of the result follows. 
\end{proof}

%%=====================
\subsubsection{Approximate case}
\lb{sssec:2approx}

We now show how to use the eigenbasis method to build approximate purifications. 

\begin{result}
\lb{lem:2approx}
Let $\rho$ denote a multipartite density matrix with spectral decomposition as in \eref{eq:eig}, and let $D=\trm{OSR}(\rho)$.
The eigenbasis method provides a density matrix $\sigma_{s}$ with $\trm{rank}(\sigma_{s})=s$ such that 
\be
\rk_{\trm{puri}}(\sigma_{s})\leq D s^{2}\, ,
\lb{eq:rpsigma2}
\ee
and its distance to $\rho$ is
\be
||\rho-\sigma_{s}||_{1}\leq 2 \sum_{i=s+1}^{n}\lambda_{i}\, .
\lb{eq:d1}
\ee
\end{result}

\begin{proof}
We construct $\sigma_{s}$ with the largest $s$ eigenvalues of $\rho$, i.e. 
\be
\sigma_{s} = \frac{1}{\mc{N}}\sum_{i=1}^{s} \lambda_{i}|\phi_{i}\ra\la\phi_{i}|\, ,
\ee
where $\mc{N}=\sum_{i=1}^{s}\lambda_{i}$.
By direct calculation one can see that \eref{eq:d1} holds with equality. Applying Result \ref{lem:2} to $\sigma_{s}$ yields equation~\eref{eq:rpsigma2}.  
\end{proof}

%Note that with this method the purification rank scales only multiplicatively in the number of eigenvalues of the truncated density matrix. 
Clearly, this method yields good approximations for rapidly decaying distributions of eigenvalues, for which the distance \eref{eq:d1} is small.
In the next section we make these statements precise, and compare this method to the sos polynomial method.

\subsection{Comparison of approximation methods}
\lb{ssec:compare}

We now compare the two approximation methods for a state  $\rho$ with $\trm{rank}(\rho)=n$ with the following eigenvalue distributions (where the eigenvalues are ordered in non-increasing magnitude).
\begin{enumerate}
\im Uniform distribution, defined as $\lambda_{j}=1/n$ for all $j$.
\lb{unif}
\im Equally spaced distribution, defined as $\lambda_{j}=j \frac{2}{n(n+1)}$ for $j=1,\ldots, n$.
\lb{eqspaced}
\im Random distribution, defined as $\lambda_{j}=\mc{N}b_{j}$, where $b_{j}$ is a random number in a fixed interval and $\mc{N}=1/\sum_{j=1}^{n}b_{j}$.
\lb{random}
\im One fixed eigenvalue and the rest equally spaced, defined as $\lambda_{1}=1/2$ and $\{\lambda_{j}=j\mc{N}\}_{j=2}^{n}$ where $\mc{N}=1/(n(n+1)-2)$. 
\lb{la1large}
\im Exponentially decaying distribution, defined as $\lambda_{j} = \mc{N} \exp(-bj)$ where $\mc{N}=(1-e)/(e^{-n}-1)$.
\lb{exp}
\een
For each distribution, each method provides a matrix $\sigma$ at distance $\epsilon$ from $\rho$, 
\be
||\rho-\sigma||_{1}\leq \epsilon\, ,
\lb{eq:epsilon}
\ee
such that 
\be
\trm{rank}_{\trm{puri}} (\sigma)\leq f(D,\epsilon,n)\, ,
\lb{eq:f}
\ee
where $D=\trm{OSR}(\rho)$. 
Our goal is to determine $f(D,\epsilon,n)$ with the sos polynomial method (Sec.~\ref{sssec:soscomp}) and the eigenbasis method (Sec.~\ref{sssec:eigcomp}).

\subsubsection{Sos polynomial method}
\lb{sssec:soscomp}

We start with the sos polynomial method, for which we first present analytical and then numerical results obtained with the SDP, both based on Result \ref{lem:1approx}.

First, the uniform distribution is the easiest for this method. 
Using the exact result (Result \ref{lem:1}) with $m=1$ we obtain
$D'=1$, since this distribution describes the maximally mixed state. 
The sos polynomial that constructs this exact purification is $p_{1}(\lambda)= 1/n$.

Second, the sos polynomial that best approximates the line $\lambda$ in the interval $[0,\lambda_{1}]$ is a good ansatz for the equally spaced and random distribution. 
Moreover, these polynomials can be chosen so that their distance to the distribution is independent of $n$, as we show next. 

\begin{result}
\lb{res:indepn}
Consider a distribution of $n$ eigenvalues whose largest eigenvalue $\lambda_{1}\sim 1/n$, such as the equally spaced or the random distribution. 
Then  there exists a sos polynomial $p_{k}$ such that $\sum_{i=1}^{n}|\lambda_{i}-p_{k}(\lambda_{i})|$ is independent of $n$.
\end{result}

\begin{proof}
Let $p_{k}$ denote the sos polynomial of degree $2(k-1)$ that best approximates the straight line $\lambda$ in the interval $[0,1]$. 
This can be rescaled to approximate the straight line in $[0,\lambda_{1}]$,
\be
p'_{k}(\lambda)=\lambda_{1} \:\: p_{k}(\lambda/\lambda_{1})\, .
\ee
Now, let $\varepsilon=\max_{\lambda\in[0,1]}|p_{k}(\lambda)-\lambda|$. Then 
$\max_{\lambda\in[0,\lambda_{1}]}|p'_{k}(\lambda)-\lambda| =\lambda_{1}\:\varepsilon$, 
which implies that
\be
\sum_{i=1}^{n}|p'_{k}(\lambda_{i})-\lambda_{i}|
\leq n \lambda_{1} \:\varepsilon\, .
\ee
If $\lambda_{1}\sim 1/n$, this upper bound is independent of $n$.
\end{proof}
For the distribution with one fixed eigenvalue and the rest equally spaced, 
it holds that $\lambda_{2}\sim 1/n$, hence the sos polynomial $q_{k}(\lambda)=p_{k-1}(\lambda)(\lambda-\lambda_{1})^{2}$, where $p_{k-1}$ is defined in the proof of Result \ref{res:indepn}, also has a distance independent of $n$. 
Denoting the distance $||\rho-\sigma_{k}||_{1}=\epsilon=g(k)$, and using Result \ref{lem:1approx}, this implies that one can upper bound the purification rank of $\sigma$ of these three distributions by a function independent of $n$, namely $\rank_{\trm{puri}}(\sigma)\leq O(D^{g^{-1}(\epsilon) -1})$. 

Finally, for the exponentially decaying distribution, we present an ansatz of sos polynomials whose distance decreases exponentially with $k$ and is independent of $n$.
\begin{result}
Consider the exponentially decaying distribution of eigenvalues  $\lambda_{j}=ae^{-bj}$ (with $a,b>0$) for $j=1,\ldots,n$, and  the sos polynomial that passes through the largest $k-2$ eigenvalues and 0, constructed as 
\be
&& p_{k}(\lambda)= \lambda^{2} \sum_{r=1}^{k-2}  \frac{1}{\lambda_{r}} l_{r}(\lambda)\\
&&l_{r}(\lambda)
=\prod_{j=1, j\neq r}^{k-2}
\left(\frac{\lambda - \lambda_{j}}{\lambda_{r}-\lambda_{j}}\right)^{2}\, .
\lb{eq:sospoly}
\ee
The distance from this polynomial to the distribution decays exponentially in $k$,
 i.e.
\be
\sum_{i=1}^{n}|\lambda_{i}-p_{k}(\lambda_{i})|  \leq O(e^{-k})\, .
\ee
\label{res:exp}
\end{result}
\begin{proof} See Appendix~\ref{app:sospoly}.
\end{proof}
The exponential decrease in the distance 
\be
||\rho-\sigma_{k}||_{1} \leq A \exp(-B k)\, ,
\lb{eq:indepn}
\ee
implies that the purification rank of $\sigma$ scales polynomially in $D$,  
\be
\trm{rank}_{\trm{puri}}(\sigma)\leq 
\frac{D^{\ln (A/\epsilon)/B} - 1}{D-1} 
=  O(D^{\ln (1/\epsilon)-1})\, .
\lb{eq:1approx-exp}
\ee
For the exponentially decaying distribution, we provide $A$ and show that $B=b$ in Appendix~\ref{app:sospoly}. 

Let us now analyze how well these eigenvalue distributions are approximated with the SDP.
For distributions \ref{eqspaced} to \ref{exp}, the SDP gives an exponential decrease of the distance in $k$ and independent of $n$ (equation~\eref{eq:indepn}) with 
$A\approx 4$ and $B\approx 2$ for distributions \ref{eqspaced}, \ref{random},
$A\approx 3$ and $B\approx 1.3$ for \ref{la1large}, 
and for  \ref{exp} with $b=1$ we find $A\approx 4$ and $B\approx 1.3$ 
(see Figs.~\ref{fig:4distr_n100_laen1_fits}, \ref{fig:4distr_indepn}).  
The sos polynomials found by the SDP for the equally spaced distribution are shown in  figure~\ref{fig:poly_eqspaced_kmax7_n100_laiovla1_4}.
We remark that the numerics does not improve beyond $k\approx4$, as very small numbers (such as powers of small eigenvalues) are numerically treated as 0. 
We have rescaled the eigenvalues $\lambda_{i} \to \lambda_{i}/\lambda_{1}$ for all $i$, 
which allows us to use the SDP until $k\approx 7$ for some distributions.

\begin{figure}[thb]\centering
\psfrag{...}{...}
\includegraphics[width=0.45\textwidth]{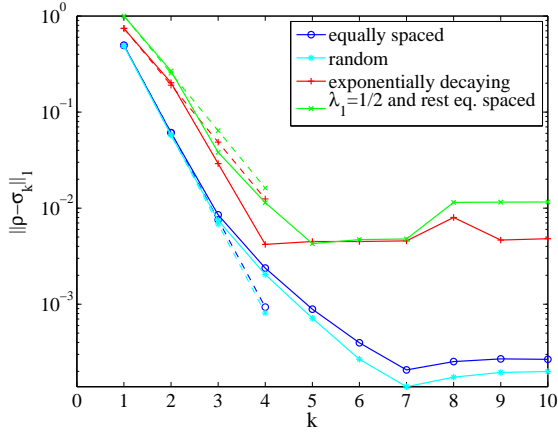}
\caption{The distance $||\rho-\sigma_{k}||_{1}$ vs.~$k$ for the distributions \ref{eqspaced} to \ref{exp} with $n=100$ as found by the SDP (solid lines) and the exponential fits (dashed lines). The figure shows an exponential decrease in the distance for small values of $k$, and the program does not reduce the distance further due to numerical errors.
}  
\label{fig:4distr_n100_laen1_fits} 
\end{figure}

\begin{figure}[thb]\centering
\includegraphics[width=0.45\textwidth]{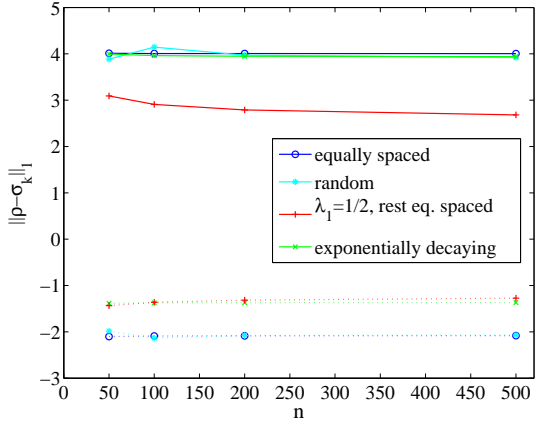}
\caption{The factors $A$ (solid line) and $B$ (dashed line) (see equation~\eref{eq:indepn}) obtained with the fits of the results of the SDP vs.~$n$. This shows that $A$ and $B$ are independent of $n$. 
}  
\label{fig:4distr_indepn} 
\end{figure}

\begin{figure}[thb]\centering
\includegraphics[width=0.45\textwidth]{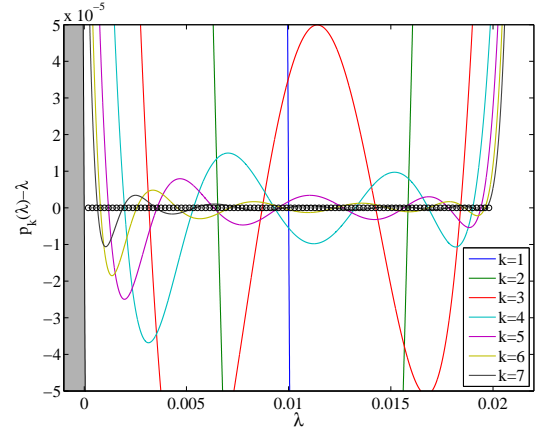}
\caption{The sos polynomial found by the SDP minus $\lambda$, $p_{k}(\lambda)-\lambda$, for various $k$, for the equally spaced distribution of eigenvalues with $n=100$. 
The black, open circles are the eigenvalues. 
The larger $k$, the smaller the distance from the polynomial to the eigenvalues.
The shaded area is that for which $p_{k}(\lambda)-\lambda \leq -\lambda$ i.e.~it is unattainable for the sos polynomial.
}  
\label{fig:poly_eqspaced_kmax7_n100_laiovla1_4} 
\end{figure}

%%%=========================
\subsubsection{Eigenbasis method}
\lb{sssec:eigcomp}

We now use the eigenbasis method (Result \ref{lem:2approx}) to upper bound the purification rank of $\sigma$ (equation~\eref{eq:f}) for the eigenvalue distributions \ref{unif} to \ref{exp}.
\ben
\im Uniform distribution.  
This is the hardest distribution, since the smallest eigenvalues are as large as possible. We obtain
\be
\trm{rank}_{ \trm{puri}}(\sigma)
&\leq& Dn^{2}(1-\epsilon/2)^{2} \, .
\lb{eq:method2-uniform}
\ee
\im Equally spaced distribution. 
We obtain
\be
\trm{rank}_{ \trm{puri}}(\sigma)& \leq& \frac{D}{4}\left(\sqrt{1+4n(n+1)(1-\epsilon/2)} - 1\right)^{2} \nn \\  
& \approx &Dn^{2}(1-\epsilon/2)  \, .
\ee
\im Random distribution. 
The distance depends on the particular random distribution;
in the worst case, it is the uniform distribution, hence it is upper bounded by equation~\eref{eq:method2-uniform}. 
\im One fixed eigenvalue and the rest equally spaced. 
We obtain
\be
\trm{rank}_{ \trm{puri}}(\sigma) & \leq&  \frac{D}{4}\left(\sqrt{1+4n(n+1)(1-\epsilon)+8\epsilon} -1 \right)^{2}   \nn\\
& \approx & Dn^{2}(1-\epsilon)\, .
\ee

\im Exponentially decaying distribution.  
Assuming that $e^{-n}\ll \epsilon$, 
the purification rank of $\sigma$ grows linearly in $D$ and quadratically in $\ln (1/\epsilon)$, i.e.
\be
\rk_{\trm{puri}}(\sigma) 
&\leq&\frac{D}{b^{2}} (\ln (2/\epsilon))^{2} \nn\\
&=& O(D \ln (1/\epsilon)^{2})\, .
\lb{eq:eigmethod-exp}
\ee
\een

In summary, for the uniform distribution, the sos polynomial is the best, as it shows exactly that  $D'=1$. 
For the equally spaced, 
random,
and one large eigenvalue and the rest equally spaced distributions, 
the sos polynomial method is better as it is independent of $n$ and scales polynomially in $D$.
Finally, for the exponentially decaying distribution, the eigenbasis method is better, as it scales linearly in $D$ (and quadratically in $\ln(1/\epsilon)$).
We thus see that the sos polynomial method is very robust, as it yields the same scaling for very different eigenvalue distributions, while the eigenbasis method works well only for rapidly decaying distributions of eigenvalues, in which case it works better than the other. 
%% 2013-10-14 Add here comment on thermal Hamiltonians

% %%%%
%For the distribution with one large eigenvalue and the rest equally spaced we observed that the polynomial makes a huge bump between $\lambda_{1}$ and $\lambda_{2}$, implying that the infinity norm cannot be used to upper bound the distance between the two.

\section{Conclusions and outlook}
\lb{sec:conclusions}

In this paper we have analyzed the efficiency of representing a mixed state as an MPDO and as a local purification, and we have shown that the latter can be arbitrarily more costly than the former.  
In particular, we have provided a family of multipartite classical states with a constant operator Schmidt rank $D$ across each linear bipartition and an unbounded purification rank $D'$ (Result \ref{countermulti}). This shows that, in the exact case, one cannot upper bound $D'$ by a function of $D$ only. 

Then we have presented two constructive purifications methods which are applicable to any multipartite density matrix. 
The exact sos polynomial method implies that $D'\leq O(D^{m-1})$, where $m$ is the number of different eigenvalues.  
Its approximate version consists of finding the sos polynomial which passes through certain points, and the optimal one can be found with an SDP (Result \ref{lem:1approx}).  
For the four tested eigenvalue distributions, this method upper bounds $D'$ by a polynomial function of $D$ which is independent of $n$, thus showing a robust and efficient behavior.

The exact eigenbasis method implies that $D'\leq Dn^{2}$, where $n$ is the number of eigenvalues. 
Its approximate version discards the smallest eigenvalues (Result \ref{lem:2approx}), and for the exponentially decaying distribution it yields $D'$ scaling linearly in $D$  (and independent of $n$).

Our inequivalence result (Result \ref{res:ineq}) implies that a single canonical form which is both efficient and locally positive semidefinite cannot exist in the exact case. Note that this is also a numerical limitation, as contracting two-dimensional (2D) PEPS requires determining the 1D density matrices which are at the boundary \cite{Ve04a}. In order to have an efficient algorithm for 2D PEPS it is thus necessary to use an efficient description of such 1D mixed states, hence to work with the MPDO form. On the other hand, for the algorithm to be stable, it is necessary that the positivity of the operator can be checked efficiently, which means for example, that it can be checked locally. While in practice one can work with MPDOs of low dimension, one would hope to reexpress such MPDOs in terms of a purification fulfilling the latter requirement, but without increasing significantly the bond dimension. Our results show that this may not be possible. 

At the same time, the results of section~\ref{ssec:compare} show that one can construct efficient approximate purifications of various relevant eigenvalue distributions. One should nonetheless analyze how successive truncations of this approximate purification affect the total error.

%This also has numerical implications, as a 1D mixed state can be seen as the boundary theory of a 2D pure state  \cite{Ci11}. If the 2D PEPS is the ground state of a gapped system, the boundary is the thermal state of a quasilocal Hamiltonian, whose eigenvalues decay exponentially, hence admit an efficient approximate purification (with the cost of \eref{eq:eigmethod-exp}). 
%On the other hand, if the bulk describes topological order, the boundary corresponds to the thermal state of a non-local Hamiltonian, *****

Let us now mention some open questions.
Concerning the inequivalence of the two forms, 
we believe that a larger separation could be obtained  with states of the form $\rho=I\otimes I - P_{1}\otimes Q_{1} -P_{2}\otimes Q_{2}$ where $P_{i},Q_{i}$ are Hermitian operators constrained by the fact $\rho \succeq 0$, but otherwise with random entries. 
Concerning the sos polynomial method, it would be interesting to find sos polynomials whose distance to the equally spaced distribution decreases exponentially with the degree $k$. 
It would also be appealing to combine both purification methods in a single one with the best of each, 
 but this requires to split the density matrix into different `eigenspace sectors' (such as in 
$\rho=\rho_{1}+\rho_{2}$ where 
 $\rho_{1,2}=\sum_{i\leq r, i>r}\lambda_{i}|\phi_{i}\ra \la \phi_{i}|$), 
 and it is not clear how the operator Schmidt rank of $\rho$ relates to that of $\rho_{1}$ or $\rho_{2}$.
 On a more practical level, it would be worth devising a purification method that works sequentially and does not require to know the spectral decomposition of the density matrix.

Our results also have connections to other research areas. In particular, the results on decomposability of mixed states translate one-to-one to divisibility properties of completely positive (CP) maps via the Choi--Jamio\l kowski isomorphism. 
While the divisibility of CP maps with a fixed dimension of the intermediate space has been studied e.g.~in \cite{Wo08,Wo08b}, our approach would allow to extend this study to varying middle dimension. 
Finally, our investigations are also related to  communication complexity, since the positive semidefinite rank determines the quantum communication complexity \cite{Fi11} and the quantum correlation complexity \cite{Ja12} of the associated matrix.

%===============================
\emph{Acknowledgements.}
We thank M.~C.~Ba\~nuls, F.~Pastawski,  M.~Piani  for discussions.
GDLC and NS acknowledge support from the Alexander von Humboldt foundation. 
DPG acknowledges support from Spanish grants
MTM2011-26912 and QUITEMAD, and European CHIST-ERA project CQC.
JIC acknowledges support from the 
EU Integrated Project SIQS. 
%Fundaci\'o Catalunya - La Pedrera
%and Fundaci\'o CELLEX.

\bibliographystyle{apsrev}

\bibliography{/Users/gemmadelascuevas/Dropbox/Gemma/Special-files/all-my-bibliography.bib}
%===============================
\begin{widetext}
\appendix

\section{Semidefinite program}
\label{app:SDP}

Here we give an exact formulation of optimization problem of \eref{eq:sdp} as an SDP. 
As mentioned in the text, we first make the objective function linear by introducing slack variables $\{z_{i}\geq 0\}$ which upper bound the absolute values, as in equation~\eref{eq:sdp2}. 
Then we rewrite the constraints so that the variables $(z,R)$ appear on one side of the inequality and the data $\lambda$ on the other, 
\be\lb{eq:sdp3}
&&\min \: \sum_{i=1}^{d^{N}}z_{i}\\
&&\trm{s.t. } -z_{i} - \la v_k^i|R_{k}|v_k^i\ra \leq -\lambda_{i}  \qquad i=1,\ldots,d^{N} \nn\\
&&\qquad    -z_{i} +\la v_k^i|R_{k}|v_k^i\ra \leq \lambda_{i}    \qquad\quad i=1,\ldots,d^{N}   \nn\\
&&\qquad R_{k}\succeq 0\, .\nn
\ee
It is now straightforward to verify that this is an SDP optimization problem, whose standard formulation is given by
\be
\lb{eq:sdpstandard}
&&\min_{X}\: \la C,X\ra\nn\\
&&\trm{s.t. } \la A_{j},X\ra\leq b_{j} \qquad j=1,\ldots,s \\
&&\qquad X\succeq 0\, .\nn
\ee
Comparing this with \eref{eq:sdp3}, we identify that variables take the following values in our problem. 
First,  
\be
X=\textrm{diag}(z_{1},\ldots,z_{d^{N}}) \oplus R\, ,
\ee
where $\oplus$ denotes direct sum. Thus  $X$ is a positive semidefinite matrix (of size $d^{N}+k$) because the variables $z$ are nonnegative and $R$ is positive semidefinite.
Then, 
\be
C= I_{d^{N}} \oplus 0_{k}
\ee
where $I_{d^{N}}$ is the identity matrix of size $d^{N}$ and  $0_{k}$ the zero matrix of size $k\times k$.
The matrix constraints are given by
\be
&&A_{j}= \diag(0,\ldots,-1,\ldots,0) \oplus (-|v_{k}^{j}\ra \la v_{k}^{j}|) \quad 1\leq j\leq d^{N}\\
&&A_{j}= \diag(0,\ldots,-1,\ldots,0) \oplus (|v_{k}^{j-d^{N}}\ra \la v_{k}^{j-d^{N}}|) \quad d^{N}<j\leq 2d^{N}
\ee
where the $-1$ is in the $j$th position. Thus we have $s=2d^{N}$ constraints. 
Note that one can always write the constraints of \eref{eq:sdpstandard} with equalities by introducing additional slack variables \cite{Va96}. 
Finally, 
\be
b=(-\lambda_{1},\ldots,-\lambda_{d^{N}},\lambda_{1},\ldots,\lambda_{d^{N}})^{T}\, .
\ee

We remark that the SDP is strongly feasible for $k<m$. To see that, note that the vectors  $\{|v_{k}^{i}\ra\}_{i=1}^{k}$ are linearly independent, hence there exists a biorthogonal basis $\{|w_{k}^{i}\ra\}_{i=1}^{k}$. We use them to define $R=\sum_{i=1}^{k}|w_{k}^{i}\ra\la w_{k}^{i}|$. 
This $R$ is positive definite ($R\succ 0$) and we choose  $z$ to be strictly positive ($z>0$). 
This point is in the interior of the feasible region. It follows that there exists no duality gap in the SDP \cite{Va96}.

We have implemented the SDP with SeDuMi 1.3 \cite{St99} and the add-on Yalmip \cite{Lo04}, with default options.

%%============================
\section{Sos polynomials for the exponentially decaying distribution of eigenvalues}
\label{app:sospoly}

Here we prove Result \ref{res:exp}.
That is, we show that the distance from the sos polynomial of equation~\eref{eq:sospoly} (which passes through the largest $k-2$ eigenvalues and 0) 
to the exponentially decaying distribution $\lambda_{j}=a e^{-bj}$ (with $a,b>0$) 
decays exponentially with $k$, 
\be
\sum_{i=1}^{n}|\lambda_{i}-p_{k}(\lambda_{i})|  \leq O(e^{-k})\, .
\lb{eq:dist1}
\ee
We first upper bound the distance using the triangle inequality
\be
\sum_{i=1}^{n}|\lambda_{i}-p_{k}(\lambda_{i})| =
\sum_{i=k-1}^{n}|\lambda_{i}-p_{k}(\lambda_{i})|
\leq  \sum_{i=k-1}^{n} \lambda_{i}
+\sum_{i=k-1}^{n} p_{k}(\lambda_{i}) \, .
\lb{eq:triangle}
\ee
The first term in the last expression is equal to 
\be
\frac{e^{-b(k-1)}-e^{-b(n+1)}}{1-e^{-b}}\, ,
\ee 
and thus  it decreases exponentially with $k$.
In the remaining of the appendix, we will show that the last term also decreases exponentially with $k$.
 
First note that the last term of \eref{eq:triangle} can be upper bounded by
\be
\sum_{i=k-1}^{n} p_{k}(\lambda_{i}) \leq 
\sum_{i=k-1}^{n} \lambda_{i}^{2}  \: 
\sum_{r=1}^{k-2} \frac{l_{r}(0)}{\lambda_{r}} \, ,
\ee
where the first sum can again be computed exactly
\be
\sum_{i=k-1}^{n} \lambda_{i}^{2}=
a^{2}\frac{e^{-2b(k-1)}-e^{-2b(n+1)}}{1-e^{-2b}} \, .
\lb{eq:sumlai}
\ee
Now we consider $l_{r}(0)$, 
\be
l_{r}(0)=  \prod_{j=1,j\neq r}^{k-2} \frac{1}{(e^{-b(r-j)}-1)^{2}} \, ,
\ee
and split the product into a term with $j<r$ times a  term with $j>r$. 
To upper bound  the term with $j<r$, we lower bound its denominator as
\be
\prod_{j=1}^{r-1} (1-e^{-b(r-j)})^{2}
&\geq &\prod_{j=-\infty}^{r-1} (1-e^{-b(r-j)})^{2} \\
&= &\left[\exp\left(\sum_{x=1}^{\infty} \ln(1-e^{-bx})\right)\right]^{2}\, ,
\lb{eq:1den}
\ee
where we have defined the variable $x=r-j$. 
Now, note that $0\geq \ln(1-e^{-bx})\geq\ln (1-e^{-b})$. 
We lower bound this function by a straight line of the variable $e^{-bx}$, i.e. $\ln(1-e^{-bx}) \geq -\alpha e^{-bx}$, where $\alpha$ is the slope of the function. 
We determine  $\alpha$ by imposing  $ \ln(1-e^{-b})=-\alpha e^{-b}$, thus yielding
% $\alpha= -e^{b}\ln(1-e^{-b})$,
\be
\left[\exp\left(\sum_{x=1}^{\infty} \ln(1-e^{-bx})\right)\right]^{2}
&\geq&
\left[\exp\left(-\alpha \sum_{x=1}^{\infty} e^{-bx}\right) \right]^{2} \\
&=&(1-e^{-b})^{2/(1-e^{-b})} =: C\, .
\ee
(Alternatively, this can be lower bounded by noting that the leftmost term is the Pochhammer function $((e^{-b};e^{-b})_{\infty}) ^{2}$, where $(a;q)_{n}$ is defined as $(a;q)_{n}=\prod_{x=0}^{n}(1-aq^{x})$. The function $f(b)=((e^{-b};e^{-b})_{\infty})^{2}$ is lower bounded by a finite value if $b$ is larger than 0).
%takes values $f(0<b\lesssim .4) \approx 0$, and $f(1)\approx 0.25$, and it fastly grows to 1 asymptotically with growing $b$. 

Now we consider the term with  $j>r$ and lower bound its denominator as
\be
\prod_{j=r+1}^{k-2} (e^{-b(r-j)}-1)^{2}
\geq 
\prod_{j=r+1}^{j^{\star}}(1-e^{-b(r-j)})^{2} =: D\, ,
\lb{eq:2den}
\ee
where $j^{\star}$ is the smallest $j$ such that $e^{-b(r-j)}-1\geq 1$,  that is, $j^{\star} = r +\lceil(\ln  2) / b\rceil  $ (for example, for $b\geq 1$, $j^{\star}=r+1$).
Note that $C$ and $D$ are independent of $k$, and the number of terms in $D$ is independent of $r$ (it only depends on $b$).

Using the lower bounds for the denominators of the parts with $j<r$ and that with $j>r$, we find 
\be
l_{r}(0)
\leq 
\frac{1}{CD}\,.
\ee
We are interested in the sum of $l_{r}(0)/\lambda_{r}$, which is
\be
\sum_{r=1}^{k-2} \frac{l_{r}(0)}{\lambda_{r}} = 
(e^{b(k-1)}-e^{b})\:
\frac{1}{(e^{b}-1)aCD}\, .
\lb{eq:sumL0}
\ee
Finally, putting together \eref{eq:sumlai} and \eref{eq:sumL0} we find 
\be
\sum_{i=k}^{n} p_{k}(\lambda_{i}) 
&\leq &
(e^{-2b(k-1)}-e^{-2b(n+1)}) \:
(e^{b(k-1)}-e^{b})\:
\frac{a}{(1-e^{-2b})(e^{b}-1)CD} \nn\\
&=&Ae^{-bk} + O(1)\, ,
\ee
where we have defined the constant $A$ as the prefactor of $e^{-bk}$ and the rest as independent of $k$. This proves the claim \eref{eq:dist1} and thus Result \ref{res:exp}.

\end{widetext}

\end{document}